\documentclass[aps,prl,twocolumn,showpacs,amsmath,amssymb,10pt]{revtex4-1}

\usepackage{amsmath}
\usepackage{amssymb}
\usepackage{latexsym}
\usepackage{amsfonts}
\usepackage{epsfig}
\usepackage{psfrag}
\usepackage{graphicx}
\usepackage{color}
\usepackage{hyperref}

\usepackage{graphicx}

\usepackage{amsmath}

\newcommand{\ket}[1]{\left|#1\right>}
\newcommand{\bra}[1]{\left<#1\right|}   
\newcommand{\braket}[2]{\left<#1|#2\right>}
\newcommand{\nn}{\nonumber\\}

\newcommand{\f}[1]{\mbox{\boldmath$#1$}}
\newcommand{\fk}[1]{\mbox{\boldmath$\scriptstyle#1$}}

\newcommand{\bea}{\begin{eqnarray}}
\newcommand{\ea}{\end{eqnarray}}
\newcommand{\eea}{\end{eqnarray}}
\newcommand{\ord}{\,{\cal O}}

\begin{document}

\title{Strong Nernst-Ettingshausen effect in folded graphene}

\author{Friedemann Queisser and Ralf Sch\"utzhold}

\email{ralf.schuetzhold@uni-due.de}

\affiliation{
Fakult\"at f\"ur Physik, Universit\"at Duisburg-Essen, 
Lotharstrasse 1, 47057 Duisburg, Germany}

\date{\today}

\begin{abstract}
We study electronic transport in graphene under the influence of a 
transversal magnetic field $\f{B}(\f{r})=B(x)\f{e}_z$ with the asymptotics 
$B(x\to\pm\infty)=\pm B_0$, which could be realized via a folded graphene 
sheet in a constant magnetic field, for example. 
By solving the effective Dirac equation, we find robust modes with a finite 
energy gap which propagate along the fold -- where particles and holes move 
in opposite directions. 
Exciting these particle-hole pairs with incident photons would then generate 
a nearly perfect charge separation and thus a strong magneto-thermoelectric 
(Nernst-Ettingshausen) or magneto-photoelectric effect 
-- even at room temperature.  
\end{abstract}


\maketitle

\paragraph{Introduction}  

The Nernst-Ettingshausen effect \cite{Nernst-Ettingshausen} 
describes the generation of an 
electric current (or voltage) by a temperature gradient in the presence 
of a magnetic field. 
Such  thermoelectric effects facilitate the direct conversion of thermal 
into electric energy and thus are of general interest.  
Obviously, the $\mathfrak C$ (charge), $\mathfrak P$ (parity), and 
$\mathfrak T$ (time reversal) symmetries must be broken for such an 
effect to occur. 
One way to achieve this is a magnetic field in a suitable geometry: 
trajectories of opposite charge carriers are bent to antipodal directions. 
However, the mean free path in usual materials is too short to generate an 
efficient charge separation in that way -- at least at room temperature. 
For example, the classical cyclotron radius $r=m_ev/(q_eB)$ of a free 
electron at room temperature in a magnetic field $B$ of one Tesla 
$r=\ord(\mu\rm m)$ is much larger than the typical mean free path 
(in the nanometer range). 
Thus, the Nernst-Ettingshausen effect is strongly suppressed by multiple 
scattering events and dissipation etc.

This motivates the study of graphene 
\cite{Wallace,roomtemperature,velocity,Exp-paper,Reviews}, 
since this system offers a comparably long mean free path 
and a large electron mobility, a linear (pseudo-relativistic) 
dispersion relation at low energies (i.e., near the Dirac points), 
and a very large Fermi velocity 
$v_{\rm F}\approx10^6\rm m/s$ \cite{velocity},
see also \cite{Nernst-in-Graphene-th,Nernst-in-Graphene-exp}. 
In this case, the pseudo-relativistic cyclotron radius at room temperature 
in a magnetic field of one Tesla is much smaller (some tens of nanometers). 
In this regime, quantum effects should be taken into account -- 
even at room temperature \cite{roomtemperature}. 

\begin{figure}[h]
\includegraphics[width=.7\columnwidth]{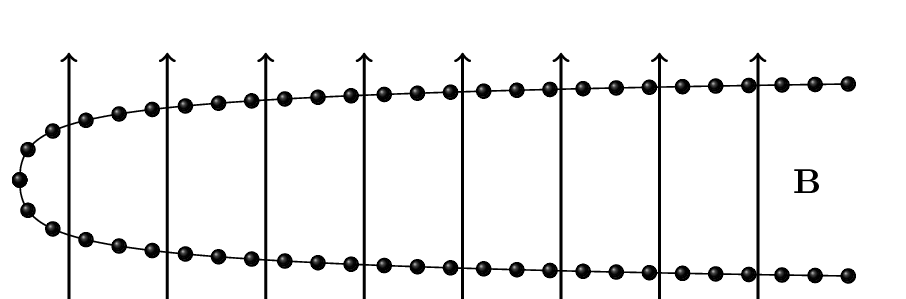}
\caption{Sketch of the considered set-up.}
\label{figure-folding}
\end{figure}

In the following, we consider folded graphene in a transversal magnetic 
field, see Fig.~\ref{figure-folding}. 
In principle, the folding of graphene has already been realized 
experimentally, see, e.g., \cite{folded-th,folded-exp}. 
This set-up is advantageous since we avoid real edges in graphene which 
are typically not perfect and contain cracks or other defects which might 
induce scattering, coupling to vibrational degrees of freedom, or further 
unwanted effects.  
Form a theoretical point of view, these edges can only be described in 
idealized cases, e.g., via effective boundary conditions which then depend 
on the concrete realization (e.g., zigzag or armchair structure 
\cite{edgestates,edge-magnetic-field-exp,edge-magnetic-field-th}).

\paragraph{Eigen-modes}  

We consider length scales (e.g., curvature radius of fold) far above 
the lattice spacing of graphene $\approx0.25~\rm nm$ and energies of 
1~eV or below.  
In this limit, we may describe the low-energy behavior by an 
effective Dirac equation in 2+1 dimensions $(\hbar=q_e=1)$ 
\begin{eqnarray}
\label{Dirac}
i\gamma^\mu\left(\partial_\mu+iA_\mu\right)\Psi=0 
\,,
\end{eqnarray}
with $x^\mu=[v_{\rm F}t,x,y]$, where $v_{\rm F}\approx10^6\rm m/s$ 
is the Fermi velocity \cite{Dirac}. 
The Dirac matrices $\gamma^\mu=[\sigma^z,i\sigma^y,-i\sigma^x]$ 
acting on $\Psi=[\psi_1,\psi_2]$
are related to the Pauli matrices $\sigma^{x,y,z}$.
In the Landau gauge, the vector potential $A_\mu=[0,0,A(x)]$ 
generates the magnetic field $B(x)=\partial_xA(x)$ with the asymptotics 
$B(x\to\pm\infty)=\pm B_0$.

In view of the translation symmetry in $t$ and $y$, we can make the 
separation ansatz for the modes 
\bea
\Psi(t,x,y)=\exp\left\{-iEt+iky\right\}\,\Psi^{E,k}(x)
\,,
\ea
arriving at the two coupled equations 
\begin{eqnarray}
\label{Dirac-coupled}
iv_\mathrm{F}[\partial_x+k+A(x)]\psi^{E,k}_2(x)&=& E\psi^{E,k}_1(x)
\nn
iv_\mathrm{F}[\partial_x-k-A(x)]\psi^{E,k}_1(x)&=& E\psi^{E,k}_2(x)
\,.
\end{eqnarray}
Hence, we can choose $\psi^{E,k}_1(x)$ to be real, for example, 
while $\psi^{E,k}_2(x)$ is imaginary.  
We observe a particle-hole symmetry since replacing $E\to-E$ and 
$\psi^{E,k}_2\to-\psi^{E,k}_2$ yields a new solution 
$\Psi^{-E,k}=\sigma^z\Psi^{E,k}=(\Psi^{E,k})^*$.

The two first-order equations~(\ref{Dirac-coupled}) can be combined 
into one second-order equation 
\bea
v^2_\mathrm{F}[k+A(x)+\partial_x][k+A(x)-\partial_x]\psi^{E,k}_1
=E^2\psi^{E,k}_1
\,,
\ea
and analogously for $\psi^{E,k}_2$ with 
$\partial_x\leftrightarrow-\partial_x$.
This equation can be cast into the form of a one-dimensional Schr\"odinger 
equation ${\cal H}_k\psi^{E,k}_1=E^2\psi^{E,k}_1$ with the Hamiltonian 
${\cal H}_k=v^2_\mathrm{F}(-\partial_x^2+{\cal V}_k)$ containing the 
effective potential ${\cal V}_k=[k+A(x)]^2+A'(x)$.
Since this Hamiltonian is self-adjoint ${\cal H}_k={\cal H}_k^\dagger$ 
and the potential ${\cal V}_k$ has the asymptotics 
${\cal V}_k(x\to\pm\infty)=\infty$, we get a complete set of discrete, 
orthonormal, and localized (in $x$) eigen-functions $\psi^{E,k}_1(x)$ 
for every value of $k$.
These modes are non-degenerate for each $k$, i.e., the energy bands 
$E(k)$ do not cross \cite{CourantHilbert}. 
Due to the particle-hole symmetry, each of these eigen-functions 
$\psi^{E,k}_1(x)$ corresponds to a pair of modes $\Psi^{\pm E,k}(x)$ 
of the original problem (\ref{Dirac-coupled}) with opposite energies. 
Furthermore, with ${\cal D}_k=k+A(x)-\partial_x$, we may write 
${\cal H}_k=v^2_\mathrm{F}{\cal D}_k^\dagger{\cal D}_k$ which shows that 
${\cal H}_k$ is non-negative (and thus $E$ is real). 
In addition, ${\cal H}_k$ cannot have a zero eigen-value $E=0$ since the 
corresponding $\psi^{E=0,k}_1(x)$ must satisfy ${\cal D}_k\psi^{E=0,k}_1=0$, 
which gives  
$\psi^{E=0,k}_1(x)\propto\exp\{kx+\int dx\,A(x)\}$ 
and analogously for $\psi^{E=0,k}_2(x)$. 
Due to the asymptotics $B(x\to\pm\infty)=\pm B_0$ and thus 
$A(x\to\pm\infty)\sim B_0|x|$, this solution is not normalizable 
and thus ${\cal H}_k$ is strictly positive for any $k$. 
Ergo, the modes do always have a finite energy gap $E\neq0$. 

\paragraph{Current}  

The current density of the modes reads 
\begin{eqnarray}
\label{current-density}
j^\mu_{E,k}
=
v_\mathrm{F}\overline{\Psi}_{E,k}\gamma^\mu\Psi_{E,k}
=
v_\mathrm{F}\Psi^\dagger_{E,k}\gamma^0\gamma^\mu\Psi_{E,k}
\,. 
\end{eqnarray}
The zeroth component $j^0=v_\mathrm{F}\rho$ is simply given by the 
density $\rho=|\psi_1^{E,k}|^2+|\psi_2^{E,k}|^2$. 
As one would expect, $j^x$ vanishes identically since $\psi^{E,k}_1(x)$ 
is real and $\psi^{E,k}_2(x)$ imaginary, cf.~Eq.~(\ref{Dirac-coupled}).
Using the same argument, the current density in $y$-direction 
simplifies to 
\begin{eqnarray}
j^y=iv_\mathrm{F}
\left(\psi_2^{E,k}\right)^*\psi_1^{E,k}
-{\rm h.c.}
=
-2iv_\mathrm{F}\psi_1^{E,k}\psi_2^{E,k}
\,. 
\end{eqnarray}
From the triangle inequality ($2|ab|\leq|a^2|+|b^2|$), we may infer 
$|j^y|\leq v_\mathrm{F}\rho$, i.e., the speed of the associated charge 
carriers is at most the Fermi velocity $v_\mathrm{F}$
(as expected). 

The total current in $y$-direction can be obtained by 
\begin{eqnarray}
\label{J^y}
J^y
=
\int dx\,j^y
=
-\frac{2v_\mathrm{F}^2}{E}
\int dx\,\psi_1^{E,k}[k+A(x)]\psi^{E,k}_1
\,, 
\end{eqnarray}
where we have used $v_\mathrm{F}{\cal D}_k\psi^{E,k}_1=iE\psi_2^{E,k}$ 
from Eq.~(\ref{Dirac-coupled}).
For the lowest $E^2$ modes (for a given $k$), i.e., 
the upper-most negative mode and the lower-most positive mode, 
the wave-function $\psi_1^{E,k}(x)$ corresponds to the ground state of 
${\cal H}_k$ and hence it is non-zero for all $x$ 
(node theorem \cite{CourantHilbert}).  
Since one can repeat the same line of argument for $\psi_2^{E,k}(x)$, 
the integrand $j^y=-2iv_\mathrm{F}\psi_1^{E,k}\psi_2^{E,k}$ is non-zero 
for all $x$ and hence the current $J^y$ is finite. 
But other modes could have $J^y=0$ at some $k$-value. 
However, for large enough $k>-A_{\rm min}=-{\rm min}\{A(x)\}$,  
the integrand in the above equation $\psi_1^{E,k}[k+A(x)]\psi^{E,k}_1$ 
is positive for all $x$ and thus the current has a finite value.

Furthermore, the current $J^y$ is related to the slope $dE/dk$ of the 
dispersion relation, i.e., the group velocity: 
Writing Eq.~(\ref{Dirac-coupled}) as 
$\hat H_{E,k}\ket{\Psi_{E,k}}=E\ket{\Psi_{E,k}}$, we find 
\bea
J^y
=
-\bra{\Psi_{E,k}}\frac{d\hat H_{E,k}}{dk}\ket{\Psi_{E,k}}
=
-\frac{dE}{dk}
\,,
\ea
where we have used the normalization $\braket{\Psi_{E,k}}{\Psi_{E,k}}=1$. 
Together with Eq.~(\ref{J^y}) we find that particles with $E>0$ and holes 
with $E<0$ have the opposite current (and group velocity), i.e., 
all particles (with $k>-A_{\rm min}$) move to the right and all holes move 
to the left.  
In this way, one obtains a (nearly) perfect charge separation.  

\paragraph{Asymptotics}  

It is illustrative to study the two limiting cases $k\to\pm\infty$. 
For large and positive $k$, the potential ${\cal V}_k$ can be approximated 
by ${\cal V}_k\approx k^2+2kA(x)$.  
Thus, to lowest order in $k$, we obtain $E\approx\pm v_\mathrm{F}k$, 
i.e., these modes propagate with a speed close to the Fermi velocity. 
Going to the next order in $k$, we may expand $A(x)$ around its minimum 
at $x_0$ where the magnetic field $B(x_0)=0$ vanishes 
$A(x)\approx A_{\rm min}+A''(x_0)(x-x_0)^2/2$ 
and obtain harmonic oscillator eigen-functions centered at $x_0$
[assuming that $A''(x_0)\neq0$]. 
Since the stiffness of the potential behaves as $kB'(x_0)$, the modes 
are strongly localized around $x_0$ for large $k$ and basically propagate 
along the $x_0$-line where the magnetic field vanishes. 
For fixed and large $k$, these modes have equidistant values of $E$
where the distance scales with $\sqrt{B'(x_0)}$. 

For large and negative $k$-values, the minima of the potential ${\cal V}_k$ 
are given by $A(x_\pm)+k=0$ and thus the modes are localized at large and 
nearly opposite values of $x_\pm\sim\pm|k/B_0|$ due to 
$A(x\to\pm\infty)\sim B_0|x|$.
In this regime, $A(x)$ is approximately linear and thus we recover the 
harmonic oscillator eigen-functions corresponding to the usual 
(pseudo-relativistic) Landau levels in a constant magnetic field 
\cite{graphene-Hall}.
Note, however, that the eigen-functions $\psi^{E,k}_1(x)$ are linear 
superpositions of the Landau levels centered at $x_+$ and $x_-$ 
with the same energy $E$.
In this limit, the eigen-energies $E$ do not depend on $k$ anymore 
($E_L^n=\pm v_\mathrm{F}\sqrt{2B_0n}$ with $n\in\mathbb N$) 
and thus the current $J^y$ also vanishes. 
Hence these modes are not so interesting for our purpose.  

\paragraph{Matrix elements}  

Now we are in the position to study the excitation of particle-hole pairs by 
incident photons (in the infra-red or optical regime). 
In second quantization, the interaction Hamiltonian reads 
\begin{eqnarray}
\hat H_{\rm int}
=
\int dx\,dy\;\hat{\overline{\Psi}}\gamma^\mu\hat A_\mu\hat{\Psi}
\,.
\end{eqnarray}
where the photon field operator $\hat A_\mu$ contains the creation 
and annihilation operators 
$\hat a_{\omega,\fk{k},\sigma}^\dagger$ and $\hat a_{\omega,\fk{k},\sigma}$
for frequency $\omega$, wavenumber $\f{k}$, and polarization $\sigma$.
The Dirac field operator $\hat{\Psi}$ is a linear combination of the 
annihilation operators for particles $\hat c_{E>0,k}\Psi_{E>0,k}$ and 
the creation operators for holes $\hat c_{E'<0,k'}^\dagger\Psi_{E'<0,k'}$.  

\begin{figure}[h]
\includegraphics[width=.9\columnwidth]{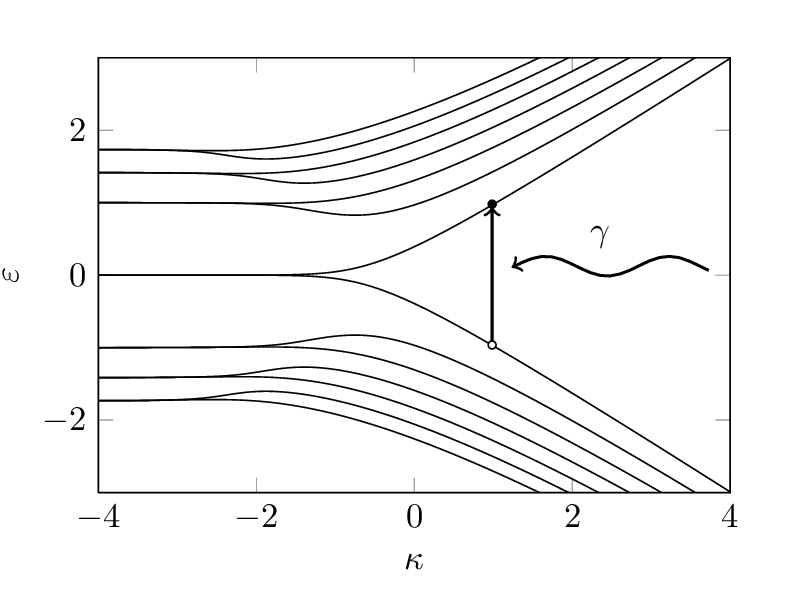}
\caption{Dispersion relation of the lowest bands with 
$\kappa=k\ell_B$ and $E=\varepsilon v_\mathrm{F}\sqrt{2B_0}$ and sketch 
of the photo-absorption.}
\label{figure-bands}
\end{figure}

If we now consider the transition matrix elements 
$\bra{{\rm out}}\hat U_{\rm int}\ket{{\rm in}}$ 
with an initial photon  
$\ket{{\rm in}}=\hat a_{\omega,\fk{k},\sigma}^\dagger\ket{0}$
and a final particle-hole pair 
$\ket{{\rm out}}=\hat c_{E>0,k}^\dagger\hat c_{E'<0,k'}^\dagger\ket{0}$,
we get to first order in perturbation theory 
\bea
{\mathfrak A}^{\omega,\fk{k},\sigma}_{E,k;E',k'}
&=&
\frac{1}{\sqrt{2\omega}}\int dt\,dx\,dy\;
\overline{\Psi}_{E,k}\gamma^\mu A_\mu^\sigma\Psi_{E',k'}\times 
\nn
&&\times 
e^{+iEt-iky}
e^{-i\omega t+i\fk{k}\cdot\fk{r}}
e^{-iE't+ik'y}
\,,
\ea
where $A_\mu^\sigma$ encodes the polarization of the photon.  
As usual, the $t$-integral gives $\delta(\omega-E+E')$, 
i.e., energy conservation. 
Since the wavelength of the photons under consideration 
(in the optical or infra-red regime) is much larger than the 
typical length scales of the electronic modes in graphene, 
we may neglect the photon wavenumber $\f{k}$.
Therefore, the $y$-integral yields $\delta(k-k')$, i.e., we excite 
particle-hole pairs with the same wavenumber $k=k'$. 
The remaining $x$-integral reads 
\bea
\label{x-integral}
{\mathfrak A}^{\omega=E-E',\fk{k}\approx0,\sigma}_{E,k;E',k'=k}
\propto
\int dx\;\overline{\Psi}_{E,k}\gamma^\mu A_\mu^\sigma\Psi_{E',k'}
\,.
\ea
Let us first assume $A_\mu^\sigma=\rm const$ and 
consider the transition between modes of the same $E^2$
(i.e., $E=-E'$), such as the upper-most negative mode (for a given $k$) 
and the lower-most positive mode, cf.~Fig.~\ref{figure-bands}.
In this case, we may use the aforementioned particle-hole symmetry 
$\Psi_{-E,k}=\sigma^z\Psi_{E,k}$ and simplify the integrand via 
$\overline{\Psi}_{E,k}\gamma^\mu\Psi_{E',k}=
\overline{\Psi}_{E,k}\gamma^\mu\sigma^z\Psi_{E,k}$. 
Inserting $\gamma^1=i\sigma^y$ and  $\gamma^2=-i\sigma^x$ and using 
the properties of the Pauli matrices, we see that the matrix element 
for the photon polarization in $x$-direction $A^x$ yields the same 
expression as in the current $J^y$, cf.~Eq.~(\ref{current-density}), 
and vice versa. 
Consequently, the matrix elements (\ref{x-integral}) vanish for the photon 
polarization in $y$-direction, but yield a non-zero contribution for the 
photon polarization in $x$-direction, at least if $k$ is large enough 
[cf.~the discussion after Eq.~(\ref{J^y})]. 
Moreover, the modes with large currents $J_y$ and thus large group 
velocities $dE/dk$ do also have large matrix elements, which enhances the 
magneto-thermoelectric or magneto-photoelectric effect we are interested in.  

\paragraph{Pseudo-parity}  

Further selection rules arise if we assume reflection symmetry $B(-x)=-B(x)$ 
and thus $A(-x)=A(x)$ which yields the additional symmetry 
\begin{eqnarray}
\psi_1^{E,k}(-x) 
=
\pm i\psi_2^{E,k}(x)
=
i{\cal P}_{E,k}\,\psi_2^{E,k}(x)
\,,
\label{pseudoparity}
\end{eqnarray}
where we call ${\cal P}_{E,k}=\pm1$ the pseudo-parity of this mode.  
Recalling the particle-hole symmetry $\Psi_{-E,k}=\sigma^z\Psi_{E,k}$,
we find ${\cal P}_{-E,k}=-{\cal P}_{E,k}$. 
The pseudo-parity of a given mode can be determined easily for large 
and positive $k$, where we have 
$i\psi_2^{E,k}\approx v_\mathrm{F}k\psi_1^{E,k}/E$ 
from Eq.~(\ref{Dirac-coupled}). 
Since the wave-function $\psi_1^{E,k}(x)$ of the lowest positive mode 
(for $k\to\infty$) corresponds to the ground state of a harmonic 
oscillator, it is Gaussian and symmetric $\psi_1^{E,k}(-x)=\psi_1^{E,k}(x)$. 
Hence this mode has an even pseudo-parity ${\cal P}_{E,k}=+1$. 
The wave-function $\psi_1^{E,k}(x)$ of the next mode corresponds to the 
first excited state of a harmonic oscillator and thus is anti-symmetric 
$\psi_1^{E,k}(-x)=-\psi_1^{E,k}(x)$, which gives an odd pseudo-parity 
${\cal P}_{E,k}=-1$ and so on. 
Together with the above result ${\cal P}_{-E,k}=-{\cal P}_{E,k}$ 
we find that, for a fixed $k$, the pseudo-parity of the modes 
alternates if we go up and down in energy. 
Assuming that the modes deform continuously if $k$ changes 
[i.e., that $A(x)$ is sufficiently well-behaved], we may deduce an 
alternating pseudo-parity for all $k$. 

Now, the integrand in the matrix elements (\ref{x-integral}) behaves as 
$\psi_1^{E,k}(x)\psi_2^{E',k}(x)\pm\psi_2^{E,k}(x)\psi_1^{E',k}(x)$
for the two photon polarizations.  
Inserting Eq.~(\ref{pseudoparity}) and integrating over $x$, 
we see that the matrix elements (\ref{x-integral}) between modes of the 
same pseudo-parity vanish for photon polarizations in $x$-direction 
whereas the transition between modes of opposite pseudo-parity is 
forbidden for the other polarization. 

Yet another set of selection rules can be obtained in the asymptotic regimes. 
For $k\to\infty$ we only get transitions between modes of opposite energies 
(due to the orthogonality of the harmonic oscillator eigen-functions).
In the opposite limit $k\to-\infty$, we recover the well-known 
\cite{graphene-Hall}
properties of the Landau levels $E_L^n=\pm v_\mathrm{F}\sqrt{2B_0n}$ 
with $n\in\mathbb N$ where we only get transitions for $n\to n\pm1$.  

\paragraph{Polarization dependence}  

So far, we have discussed the case $A_\mu^\sigma=\rm const$ in 
Eq.~(\ref{x-integral}). 
This is certainly a good approximation if the polarization of the incident 
photon points in $y$ direction, i.e., is aligned with the symmetry of our 
set-up. 
However, for the other ($x$) polarization, $A_\mu^\sigma$ in 
Eq.~(\ref{x-integral}) should be replaced by the local projection of the 
photon wave function $A_\mu^\sigma$ onto the graphene plane, i.e., 
become $x$-dependent $A_\mu^\sigma(x)$. 
The profile of $A_\mu^\sigma(x)$ then depends on the incidence angle of the 
photon. 
If the photon is incident from top, i.e., propagates parallel to the external 
magnetic field $\f{k}\|\f{B}$, the two graphene sheets (top and bottom) have 
opposite projections.
Thus $A_\mu^\sigma(x)$ is anti-symmetric $A_\mu^\sigma(-x)=-A_\mu^\sigma(x)$
and the above selection rules are reversed. 
If the photon propagates perpendicularly through the fold ($\f{k}\perp\f{B}$), 
we get a symmetric projection function $A_\mu^\sigma(-x)=A_\mu^\sigma(x)$ 
which vanishes far away from the folding region (i.e., for large $|x|$).
In this case, the above selection rules do still apply, but the matrix 
elements might be reduced a bit. 
%
%

\paragraph{Example profile}  

In order to visualize the behavior of the modes by means of a concrete 
example, let us consider a magnetic field of the following form 
\bea
B(x)=B_0\tanh(\alpha x)
\,,
\ea
where $1/\alpha$ measures the width of the fold. 
For $\alpha\to\infty$, we get a step function $B(x)=B_0{\rm sign}(x)$ 
with the vector potential $A(x)=B_0|x|$, cf.~\cite{alternatingB}.  
In this limit, the mode equation~(\ref{Dirac-coupled}) can be solved 
exactly (piecewise) in terms of parabolic cylinder functions, 
cf.~\cite{exakte-Loesungen}. 
Incidentally, the spectrum for such a step function $B(x)=B_0{\rm sign}(x)$ 
can also arise for some edge states 
\cite{edge-magnetic-field-th,edge-magnetic-field-exp}. 

However, such a step function can only be a good approximation if $k$ is 
not too large and if the curvature radius of the graphene fold is much 
smaller than the typical magnetic length scale $\ell_B=1/\sqrt{B}$.
For one Tesla, we get $\ell_B\approx26~\rm nm$ while the radius of curvature 
cannot be too small since it should be much larger than the lattice spacing 
$\approx0.25~\rm nm$.  
Thus, let us consider a finite $\alpha$ and take $\alpha=1/\ell_B$ 
as an example.  
The spectrum can then be obtained numerically and is given in 
Fig.~\ref{figure-bands}. 
The spectra for other values of $\alpha$ are qualitatively similar.
As demonstrated above, the two lowest modes are monotonically 
increasing/decreasing, whereas the higher modes can have $dE/dk=0$ 
at some small $k$-values. 
For large $|k|$, we recover the asymptotics discussed above. 

\paragraph{Conclusions}  

Via the effective Dirac equation~(\ref{Dirac}), we studied the low-energy 
behavior of electronic excitations in graphene under the influence of a 
transversal magnetic field $B(x)$ with the asymptotics 
$B(x\to\pm\infty)=\pm B_0$.
Such a field profile $B(x)$ arises within a folded graphene sheet in a 
constant magnetic field, for example, see Fig.~\ref{figure-folding}. 
Based on general arguments, we find a discrete set of modes 
(see also \cite{snake}) 
which are localized near the fold (i.e., the zero of the magnetic field) 
and propagate along it with a significant fraction of the Fermi velocity. 

Due to particle-hole symmetry, the dispersion relations $E(k)$ of these 
modes (cf.~Fig.~\ref{figure-bands}) are symmetric around the $E=0$ axis, 
but do never cross it.
Thus, these modes have a finite energy gap (for each $k$) with the 
characteristic energy scale being set by the (pseudo-relativistic) 
Landau level energy $E_L=v_\mathrm{F}\sqrt{2B_0}$.  
For a magnetic field of one Tesla, we have $E_L\approx36~\rm meV$, 
which corresponds to 400~Kelvin. 
%
%
The group velocity $dE/dk$ is related to the current $J^y$ and we find 
that particles and holes move in opposite directions. 
Apart from some minor exceptions, all particles move to the right and 
all holes move to the left -- i.e., we get a nearly perfect charge 
separation. 
In view of this pre-determined direction, the finite energy gap, 
and the fact that these localized modes are qualitatively independent 
of the shape of $A(x)$, we expect that they are quite robust against 
perturbations. 
In addition, we consider the propagation within a (curved) graphene sheet,  
i.e., far away from any edges with defects etc. 

Finally, we study the excitation of particle-hole pairs in these modes via 
incident infra-red or optical photons, i.e., the magneto-thermoelectric 
(Nernst-Ettingshausen) or magneto-photoelectric effect.  
The matrix elements display a distinct dependence on the polarization and 
the incidence angle of the photons, which should enable us to distinguish 
this effect from other phenomena experimentally. 
Furthermore, we find that those modes with comparably large group velocities 
(i.e., large currents) tend to have large matrix elements 
(at least for low-energy transitions) and thus are more strongly 
coupled to the incident photons (i.e., ``nature favors our goal''). 

\paragraph{Outlook: electric field}  

If we apply an additional electric field perpendicular to the fold and the 
magnetic field, we get an electrostatic potential 
$\Phi(x)=\beta v_\mathrm{F}A(x)$ with some constant $\beta$.
If we have $|\beta|<1$ (i.e., if the electric field is sub-critical), 
we may transform $\Phi$ away by an effective Lorentz boost in $y$-direction 
with a velocity $v_\mathrm{boost}=\beta v_\mathrm{F}$ where $v_\mathrm{F}$
plays the role of the speed of light \cite{lorentz-trafo}.  
In the Lorentz boosted frame, we get the same modes as discussed above, 
but with a reduced magnetic field $B_0'=B_0\sqrt{1-\beta^2}$. 
Since this field enters the characteristic energy scale via 
$v_\mathrm{F}\sqrt{2B_0}$, the dispersion relation after transforming back to 
laboratory coordinates reads 
\bea
E\to E'=E(1-\beta^2)^{3/4}-kv_\mathrm{F}\beta\,,
\ea
i.e., the spectrum in Fig.~\ref{figure-bands} is compressed and tilted. 

\begin{acknowledgments}
\paragraph{Acknowledgements}  
%
Fruitful discussions with A.~Lorke and M.~Schleberger 
are gratefully acknowledged. 
This work was supported by DFG (SFB-TR12). 
\end{acknowledgments}

\bibliographystyle{apsrmp4}

\begin{thebibliography}{99}

\bibitem{Nernst-Ettingshausen}
A.~von Ettingshausen, W.~Nernst, 
Annalen der Physik {\bf 265}, 343 (1886).  

\bibitem{Reviews}
M.~O.~Goerbig, Rev.~Mod.~Phys.~{\bf 83}, 1193 (2011);
A.~H.~C.~Neto, F.~Guinea, N.~M.~R.~Peres, K.~S.~Novoselov and A.~K.~Geim, 
Rev.~Mod.~Phys.\ {\bf 81}, 109 (2009);
S.~Das Sarma, S.~Adam, E.~H.~Hwang, and E.~Rossi, 
Rev.~Mod.~Phys.\ {\bf83}, 407 (2011).

\bibitem{velocity}
M.~I.~Katsnelson, I.~V.~Grigorieva, S.~V.~Dubonos, and A.~A.~Firsov, 
Nature {\bf438}, 197 (2005). 

\bibitem{roomtemperature}
K.~S.~Novoselov, A.~K.~Geim, S.~V.~Morozov, D.~Jiang, Y.~Zhang, 
S.~V.~Dubonos, I.~V.~Grigorieva, A.~A.~Firsov, 
Science {\bf 306}, 666 (2004). 

\bibitem{Exp-paper}
Y.~Zhang, Y.-W.~Tan, H.~L.~Stormer, and P.~Kim, 
Nature {\bf 438}, 201 (2005);
D.~A.~Abanin, K.~S.~Novoselov, U.~Zeitler, P.~A.~Lee, A.~K.~Geim, 
and L.~S.~Levitov, Phys.~Rev.~Lett.\ {\bf 98}, 196806 (2007). 

\bibitem{Wallace}
P.~R.~Wallace, Phys.~Rev.~{\bf 71}, 622 (1947). 

\bibitem{Nernst-in-Graphene-exp}
J.~G.~Checkelsky and N.~P.~Ong, Phys.~Rev.~B {\bf 80}, 081413(R) (2009);
Z.~Zhu, H.~Yang, Beno\^it Fauqu\'e, Y.~Kopelevich and  K.~Behnia, 
Nature Physics {\bf 6}, 26 (2010);
Y.~M.~Zuev, W.~Chang, and P.~Kim, Phys.~Rev.~Lett.~{\bf 102}, 096807 (2009);
P.~Wei, W.~Bao, Y.~Pu, C.~N.~Lau, and J.~Shi, 
Phys.~Rev.~Lett.~{\bf 102}, 166808 (2009). 

\bibitem{Nernst-in-Graphene-th}
L.~Zhu, R.~Ma1, L.~Sheng, M.~Liu, and D.~-N.~Sheng, 
Phys.~Rev.~Lett.\ {\bf 104}, 076804 (2010);
I.~A.~Luk'yanchuk, A.~A.~Varlamov, and A.~V.~Kavokin, 
Phys.~Rev.~Lett.\ {\bf107}, 016601 (2011);
E.~H.~Hwang, E.~Rossi, and S.~D.~Sarma, 
Phys.~Rev.~B {\bf 80}, 235415 (2009). 

\bibitem{folded-th}
E.~Prada, P.~San-Jose, L.~Brey, Phys.~Rev.~Lett.~{\bf105}, 106802 (2010);
D.~Rainis, F.~Taddei, M.~Polini, G.~Le\'on, F.~Guinea, and V.~I.~Fal'ko, 
Phys.~Rev.~B {\bf 83}, 165403 (2011);
N.~Yang, X.~Ni, J.-W.~Jiang, and B.~Li, 
Appl.~Phys.~Lett.\ {\bf 100}, 093107 (2012). 

\bibitem{folded-exp}
S.~Akc\"oltekin, H.~Bukowska, T.~Peters, O.~Osmani, I.~Monnet, 
I.~Alzaher, B.~Ban d'Etat, H.~Lebius, and M.~Schleberger, 
Appl.\ Phys.\ Lett.\ {\bf 98}, 103103 (2011); 
J.~Zhang, J.~Xiao, X.~Meng, C.~Monroe, Y.~Huang, and J.-M.~Zuo, 
Phys.~Rev.~Lett.~{\bf104}, 166805 (2010);
S.~Cranford, D.~Sen, and M.~J.~Buehler, 
Appl.~Phys.~Lett.\ {\bf 95}, 123121 (2009);
J.-H.~Yoo, J.~B.~In, J.~B.~Park, H.~Jeon, and C.~P.~Grigoropoulos, 
Appl.~Phys.~Lett.\ {\bf100}, 233124 (2012);
L.~Ortolani, E.~Cadelano, G.~P.~Veronese, C.~D.~E.~Boschi, 
E.~Snoeck , L.~Colombo, and V.~Morandi,
Nano Lett., {\bf 12}, 5207 (2012);
K.~Kim1, Z.~Lee, B.~D.~Malone, K.~T.~Chan, B.~Alem\'an, W.~Regan, 
W.~Gannett, M.~F.~Crommie, M.~L.~Cohen, and A.~Zettl,
Phys.~Rev.~B {\bf83}, 245433 (2011). 

\bibitem{edgestates}
M.~Fujita, K.~Wakabayashi, K.~Nakada and K.~Kusakabe, 
J.~Phys.~Soc.~Jpn.~{\bf 65} 1920 (1996);
K.~Nakada, M.~Fujita, G.~Dresselhaus and M.~S.~Dresselhaus, 
Phys.~Rev.~B {\bf 54}, 17954 (1996). 

\bibitem{edge-magnetic-field-th}
S.~Park and H.-S.~Sim, Phys.~Rev.~B {\bf 77}, 075433 (2008);
D.~A.~Abanin, P.~A.~Lee, and L.~S.~Levitov, 
Phys.~Rev.~Lett.\ {\bf 96}, 176803 (2006);
N.~M.~R.~Peres, A.~H.~Castro Neto, and F.~Guinea, 
Phys.~Rev.~B {\bf 73}, 241403(R) (2006);
K.~Wakabayashi, M.~Fujita, H.~Ajiki, and M.~Sigrist, 
Phys.~Rev.~B {\bf 59}, 8271 (1999);
S.~Wu, M.~Killi, and A.~Paramekanti, 
Phys.\ Rev.\ B {\bf 85}, 195404 (2012);
J.~A.~M.~van Ostaay, A.~R.~Akhmerov, C.~W.~J.~Beenakker, M.~Wimmer, 
Phys.~Rev.~B {\bf 84}, 195434 (2011);
N.~M.~R.~Peres, F.~Guinea, and A.~H.~Castro Neto, 
Phys.~Rev.~B {\bf 73}, 125411 (2006);
H.~A.~Fertig, L.~Brey, Phys.~Rev.~Lett.\ {\bf 97}, 116805 (2006);
D.~A.~Abanin, P.~A.~Lee, L.~S.~Levitov, 
Solid State Comm.\ {\bf 143}, 77 (2007);
P.~Delplace and G.~Montambaux, Phys.~Rev.~B {\bf82}, 205412 (2010);
I.~Romanovsky, C.~Yannouleas, and U.~Landman, 
Phys.~Rev.~B {\bf83}, 045421 (2011).

\bibitem{edge-magnetic-field-exp}
R.~Ribeiro, J.-M.~Poumirol, A.~Cresti, W.~Escoffier, M.~Goiran, 
J.-M.~Broto, S.~Roche, and B.~Raquet, 
Phys.~Rev.~Lett.\ {\bf 107}, 086601 (2011);
S.~Minke, S.~H.~Jhang, J.~Wurm, Y.~Skourski, J.~Wosnitza, C.~Strunk, 
D.~Weiss, K.~Richter, and J.~Eroms, 
Phys.\ Rev.\ B {\bf 85}, 195432 (2012). 

\bibitem{Dirac}
G.~W.~Semenoff, Phys.~Rev.~Lett.~{\bf53}, 2449 (1984). 

\bibitem{CourantHilbert}
R.~Courant and D.~Hilbert, 
\textit{Methoden der Mathematischen Physik} (Springer, Berlin, 1924). 

\bibitem{graphene-Hall}
V.~P.~Gusynin, and S.~G.~Sharapov, Phys.~Rev.~Lett.~95, 146801 (2005);
M.~L.~Sadowski, G.~Martinez, and M.~Potemski,C.~Berger and W.~A.~de Heer, 
Phys.~Rev.~Lett.~{\bf97}, 266405 (2006). 

\bibitem{alternatingB}
L.~Dell'Anna and A.~De Martino, Phys.~Rev.~B {\bf79}, 045420 (2009);
A.~De Martino, L.~Dell'Anna, and R.~Egger, 
Phys.~Rev.~Lett.\ {\bf 98}, 066802 (2007). 

\bibitem{exakte-Loesungen}
S.~Kuru, J.~Negro, and L.~M.~Nieto, 
J.\ Phys.\ Condens.~Matter {\bf 21}, 455305 (2009). 

\bibitem{lorentz-trafo}
V.~Lukose, R.~Shankar, and G.~Baskaran, 
Phys.~Rev.~Lett.\ {\bf 98}, 116802 (2007). 

\bibitem{snake}
T.~K.~Ghosh, A.~De Martino, W.~H\"ausler, L.~Dell'Anna, and R.~Egger, 
Phys.~Rev.~B {\bf 77}, 081404(R) (2008);
L.~Oroszl\'any, P.~Rakyta, A.~Korm\'anyos, C.~J.~Lambert, and J.~Cserti, 
Phys.~Rev.~B {\bf 77}, 081403(R) (2008);
J.~R.~Williams and C.~M.~Marcus, Phys.~Rev.~Lett.\ {\bf 107}, 046602 (2011). 

\end{thebibliography}


\end{document}